# Electron tunneling from colloidal CdSe quantum dots to ZnO nanowires studied by time-resolved luminescence and photoconductivity experiments

*Stephanie Bley[1]\*, Michael Diez[1], Friederike Albrecht[1], Sebastian Resch[2], Siegfried R. Waldvogel[2], Andreas Menzel[3], Margit Zacharias[3], Jürgen Gutowski[1], Tobias Voss[1,4]*

[1]Institute of Solid State Physics, University of Bremen, Bremen, Germany

[2] Department of Organic Chemistry, Johannes Gutenberg-University Mainz, Mainz, Germany

[3]Department of Microsystems Engineering - IMTEK, University of Freiburg, Freiburg, Germany

[4]Institute of Semiconductor Technology, Braunschweig University of Technology, Braunschweig, Germany

**Abstract**

CdSe quantum dots (QDs) with different organic linker molecules are attached to ZnO nanowires (NWs) to study the luminescence dynamics and the electron tunneling from the QDs to the nanowires in time-resolved photoluminescence (PL) and photoconductivity measurements. The PL transients of the QD luminescence indicate two different recombination channels: the direct recombination inside the QD core and the recombination via QD surface defect states. After linking the QDs to the ZnO NW surface, photo-induced electron tunneling from an excited state of the QD into the conduction band of the nanowire becomes visible by a clear decrease of the PL decay time. Efficient electron tunneling is confirmed by a strong enhancement of the photocurrent through the functionalized nanowires in which the tunneling rate can be controlled by using different organic linker molecules.

# 1. Introduction

Nowadays, a huge interest exists in the development of hybrid nanostructure devices for application[1,2] as LEDs, transistors, catalysts, and energy harvesting devices. The increasing importance of regenerative energies, especially via photovoltaic applications, results in intense research concerning the development of inexpensive and efficient optoelectronic devices in the near ultraviolet spectral region. The major goals are to achieve a high absorption in the wide UV range, a high concentration of charge carriers, and a fast and efficient conversion of the incident photon energy into a most significant electrical signal.

Electronic and optical coupling as well as electron transfer across the internal interfaces of hybrid nanostructures open possibilities to specifically tailor their microscopic transport and luminescence processes by combining the specific properties of different organic and inorganic material systems[3]. One of the most important nanostructured hybrid solar cells has been introduced as the dye-sensitized solar cell by O'Regan and Grätzel[4] in 1991. The dye with a suitable absorbance in the UV range acts as photo sensitizer to provide electrons for the $TiO_2$ electrode, the latter being a II-VI oxide with a wide band gap of 3.2 eV and a high charge mobility.

ZnO nanowires are another efficient and low-cost material to serve as the photo electrode[5]. Well-understood characteristics of ZnO are its wide band gap (3.37 eV), high electron mobility, and high absorbance in the UV range[6-9]. Because of the large surface-to-volume ratio of ZnO nanowires, their optical and electrical properties can be efficiently tailored by surface functionalization. The optimization of the sensitizer is a very crucial point for developing hybrid solar cells or photodetectors. To achieve this goal, colloidal QDs provide many advantages. They are simple to synthesize and exhibit a high quantum efficiency. Control of size and shape of these quantum dots during growth leads to a tunable band gap. QDs show large absorption coefficients and multi-exciton generation even by absorption of a single photon[10]. Surface functionalization of ZnO nanostructures by QD systems with different QD sizes and surface modifications has been demonstrated[11,12] to lead to hybrid solar cells with high absorption over a wide spectral range. However, open questions concern the understanding of the charge carrier dynamics between sensitizer and electrode. In this respect, ZnO nanowires (NWs) functionalized with colloidal quantum dots (QDs) through organic linker molecules allow for a detailed investigation of the coupling of the 3D (NW) and 0D (QD) electronic systems. A tight and selective binding of the QDs to the ZnO NW surface can be achieved through a special design of the organic linker molecules, and the interaction between the 3D and 0D electronic systems can be precisely controlled by adjusting the lengths of the linker molecules and their electronic system.

In this work, colloidal CdSe QDs with different organic linker molecules are attached to ZnO nanowires to study the luminescence dynamics and the electron tunneling inside these hybrid nanostructures via time-resolved photoluminescence (PL) and photoconductivity (PC) experiments. By comparing the PL transients of QDs in solution with those of QDs linked to ZnO NWs, the photo-induced electron tunneling (PET) process between excited states of the QD and the nanowire is demonstrated and discussed in the frame of a rate equation model. It is demonstrated that the tunneling rate can be controlled by using organic linker molecules (ω-mercaptocarboxylic acids) with different linker lengths. The electron tunneling is further confirmed in photoconductivity experiments.

## 2. Experiment

### 2.1 ZnO nanostructure samples

The ZnO nanowires were fabricated by two different methods. ZnO nanowires obtained by aqueous chemical growth (ACG, hydrothermal method[13]) at a temperature of 80 °C are vertically oriented on a ZnO seed layer on top of a conductive FTO (fluorine doped tin oxide) glass as substrate. A scanning electron microscope (FEI dual beam system, model Nova 200, operating acceleration voltage 5 kV) was used for structural investigation of the morphology. In Figure 1 an SEM image of the ACG nanowires with diameters in the range of 50 - 300 nm is depicted.

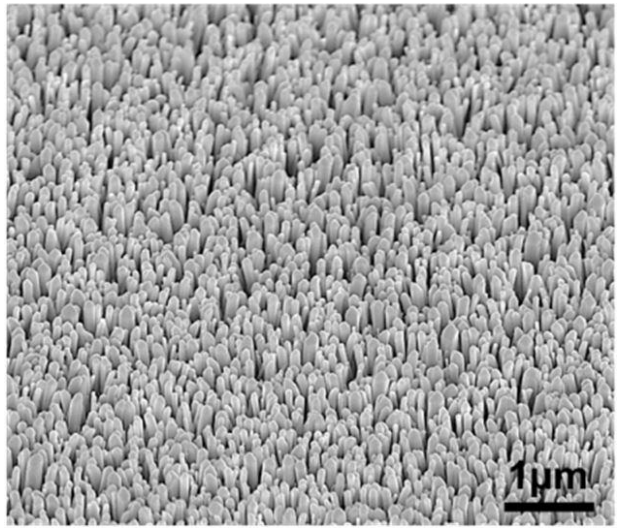

**Figure 1.** SEM images of the ACG ZnO nanowires with a diameter between 50-300 nm and a length around 1 μm.

For the studies of electrical conductivity, ZnO nanowires were fabricated in a planar geometry. They were grown on silicon substrates through a technique called atomic layer deposition (ALD) based atomic linker lithography (ASL)[14] that combines ALD with spacer lithography to obtain controlled high-aspect-ratio nanostructures on wafer scale. The nanowires were contacted with Al/Au. Figure 2 illustrates schematically such a sample together with two SEM images of the nanowires (diameter range 300 - 500 nm) and of the contacts (2.5 mm$^2$), respectively.

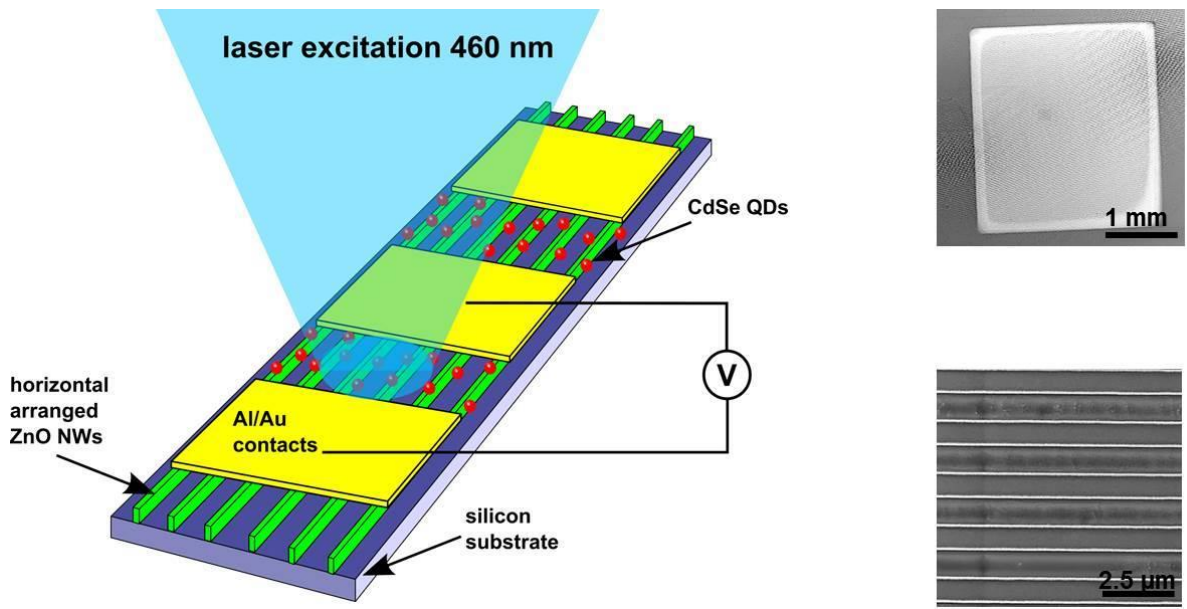

**Figure 2.** Schematic sketch of the nanowires grown on silicon substrate by a combination of ALD and ASL, including the photoconductivity measurement setup. SEM images show the Al/Au metal contact (top image) and the nanowires (bottom image).

## 2.2 Quantum dots

Colloidal CdSe QDs capped with different hydrophilic ω-mercaptocarboxylic acids (dissolved in water) were synthesized by the following methods. CdSe QDs capped with 3-mercaptopropionic acid (MPA) were prepared by a method adopted from the literature[15]. CdSe QDs capped with 6-mercaptohexanoic acid (MHA) and 11-mercaptoundecanoic acid (MUA) were obtained by a ligand exchange method[16,17] using commercially available QDs. The thiol group (SH) of mercaptocarboxylic acids passivates the QD surface. The carboxyl group (COOH) makes them soluble in water, avoids aggregation, and binds to the zinc ions at the nanowire surfaces.

Two types of commercially available QDs, i.e., CdSe QDs and CdSe/ZnS core-shell QDs, capped with hydrophobic octadecylamine (ODA) (dissolved in toluene), were additionally used. A monolayer ZnS shell results in a larger PL quantum yield[18] of the QD. Because of the higher band gap of ZnS compared to CdSe, the trapping of electrons in surface states is hindered. The amino group ($NH_2$) of octadecylamine passivates the QD surface and makes the quantum dots soluble in toluene. In Table 1, the major characteristics of the used linker molecules are summeraized, whereas in Table 2, the absorption and emission maxima of all QDs applied in solution are listed.

**Table 1**. Major characteristics of the used linker molecules.

| ligand | chemical formula | spacing moieties | property |
|---|---|---|---|
| octadecylamine (ODA) | $C_{18}H_{37}NH_2$ | 18 | hydrophobic |
| 3-mercaptopropionic acid (MPA) | $HSCH_2CH_2CO_2H$ | 2 | hydrophylic |
| 6-mercaptohexanoic acid (MHA) | $HS(CH_2)_5CO_2H$ | 5 | hydrophylic |
| 11-mercaptoundecanoic acid (MUA) | $HS(CH_2)_{10}CO_2H$ | 10 | hydrophylic |

**Table 2**. Absorption and emission maxima of all used QDs in solution.

| QD description | absorption maximum (nm) | emission maximum (nm) |
| --- | --- | --- |
| CdSe/ZnS QDs, toluene, ODA | 510 | 530 |
| CdSe QDs, toluene, ODA | 550 | 560 |
| CdSe/ZnS QDs, H$_2$O, MPA | 520 | 540 |
| CdSe/ZnS QDs, H$_2$O, MHA | 545 | - |
| CdSe/ZnS QDs, H$_2$O, MUA | 555 | - |

After ligand exchange, no photoluminescence (PL) signal was detected for MHA- and MUA-capped QDs, and the solution was only stable for two days. For this behavior, either super-saturation of the used ligands or an environment being too much acidic could be responsible reasons. In the latter case, a high density of defects is generated so that the QDs are losing their solubility.

### 2.3 QD attachment to the ZnO surface

For reference measurements of the time-resolved QD photoluminescence transients, a diluted QD solution was used. Therefore, a glass cuvette was filled with deionized water or toluene (according to the different QD types) and mixed with a pipet tip (~ µl) full of the original QD solution.

For QD coating of the ZnO surface, a pipet tip filled with the original QD solution was used to place a drop of the solution onto the nanowire surface, and the sample was dried overnight. An SEM image of the QD-coated ACG ZnO nanowires is shown in Figure 3. The QDs do not completely cover the nanowire surface but are largely concentrated in clusters at the nanowire tip.

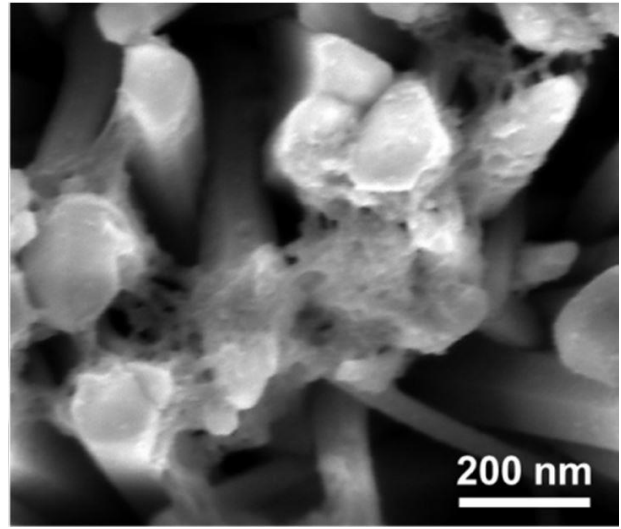

**Figure 3.** SEM image of QD-coated ACG ZnO nanowires.

In case of the contacted horizontal nanowires, a drop of the QD solution was placed between the contacts. Afterwards, the samples were carefully rinsed with deionized water to remove residual QDs.

**2.4 Optical and electrical characterization**

Optical excitation was performed with a pulsed Ti:sapphire laser (Tsunami, Spectra Physics, $\Delta t$=60 fs, f=82 MHz) in combination with a regenerative amplifier (Spitfire Pro, Spectra Physics, $\Delta t$=130 fs, f=1 kHz) and an optical parametric amplifier (Topas Prime, Spectra Physics). For time-resolved PL measurements, an excitation wavelength of 400 nm (output pulse power 6 μJ, measured at sample) was applied to achieve selective excitation of the CdSe QDs. For the detection of the time-resolved PL transients, a streak camera (model SC-10, Optronis Optoscope, Δt ≈2ps) with a scan speed of 10 ns/mm (according CCD resolution 1052 x 1400 pixel, one pixel equals 14.3 μm) was employed.

The photoconductivity measurements were performed with a Keithley SourceMeter Model 2400 with an applied constant bias voltage of 0.5 V. A sketch of the measurement procedure is shown in Figure 2. The sample was excited with laser pulses (excitation wavelength 460 nm, output pulse power 2 μJ, spot size 7 mm) between the contacts. All studies (PL and PC) were performed in air.

## 3. Results of time-resolved photoluminescence

To study the electron tunneling from the attached QDs into the ZnO NWs, time-resolved PL transients of QDs dissolved in solution have been compared with those of QDs that are attached to ZnO nanowires (Figure 4).

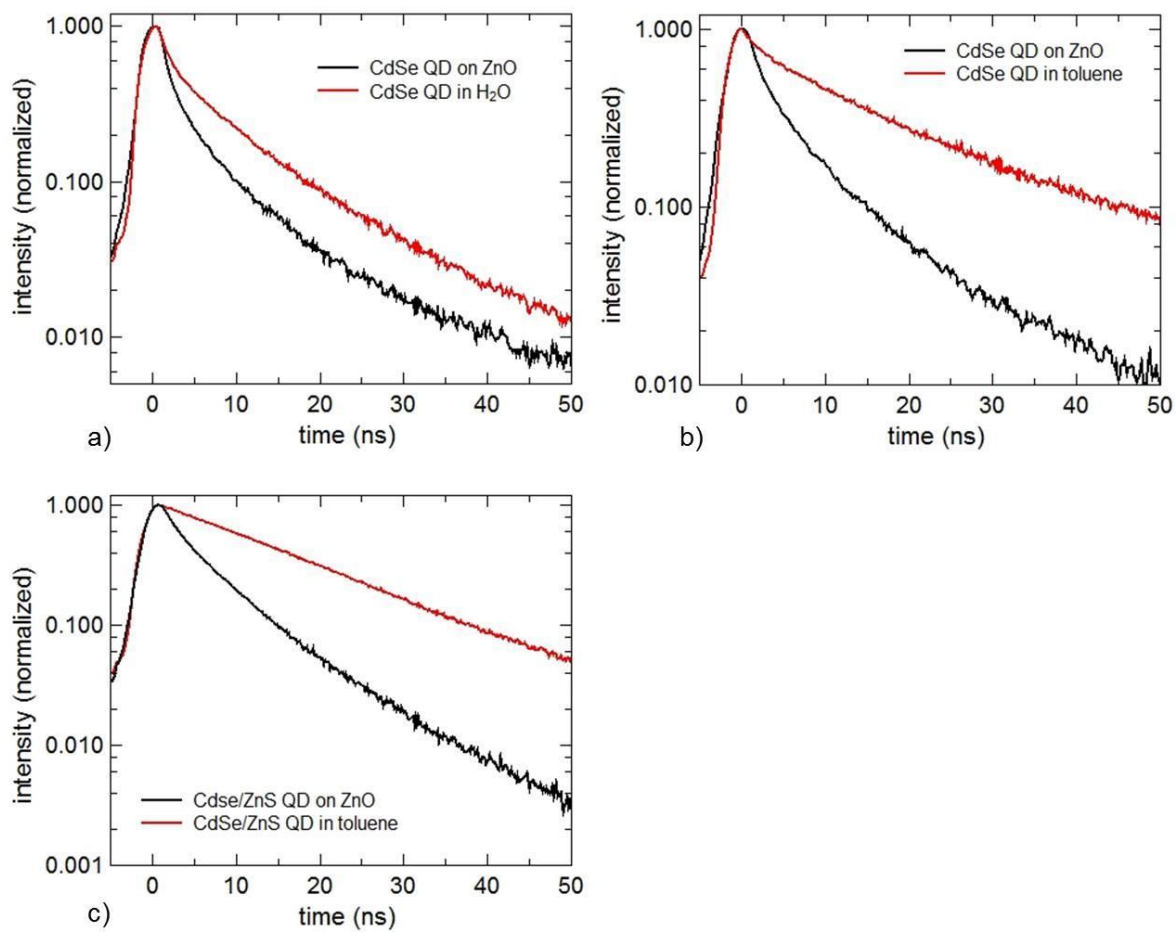

**Figure 4.** Normalized PL transients of the quantum dots. a) MPA-capped CdSe QDs in $H_2O$ (red curve) and on ACG grown ZnO NWs (black curve), b) ODA-capped CdSe QDs in toluene (red curve) and on ACG ZnO NWs, c) ODA-capped CdSe/ZnS QDs in toluene (red curve) and on ACG grown ZnO NWs.

In Figure 4(a), the time-resolved PL intensities of MPA-capped CdSe QDs dissolved in water (red) and attached to ZnO nanowires (black) are plotted. For all cases, the PL decay becomes faster when the QDs are attached to the NW. The reduction of the decay time is even more pronounced for ODA-capped CdSe QDs dissolved in toluene (red) and attached to ZnO nanowires (black) as for MPA-capped CdSe QDs in $H_2O$. A similarly pronounced decay time reduction is also found for the core-shell CdSe/ZnS QDs (Figure 4 (c)). The faster decay is particularly observable during the first 10 ns after excitation which is the time scale for direct exciton recombination[19].

To investigate the influence of the spatial separation between the QDs and the NWs on the PL decay of the QDs, PL transients of MPA- (black) and ODA-capped (red) CdSe QDs attached to ZnO nanowires are compared in Figure 5. About 2.5 ns after excitation, the PL decay becomes faster for the MPA-capped QDs than for the ODA ones.

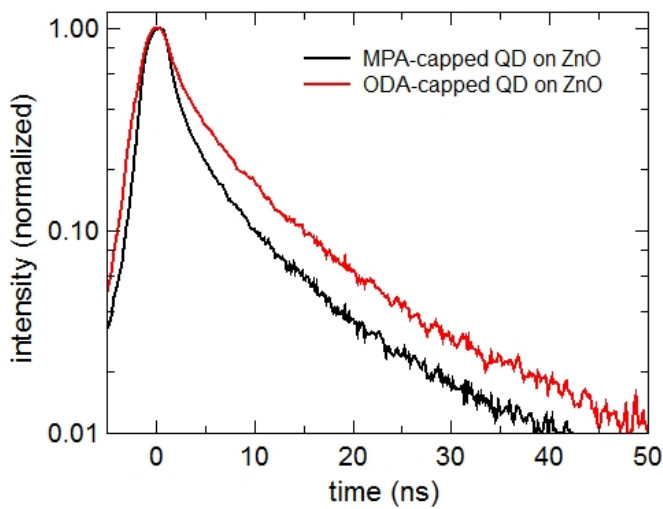

**Figure 5.** Comparison of the nanosecond-resolved transients for the QDs with different surface functionalization attached to ACG ZnO nanowires.

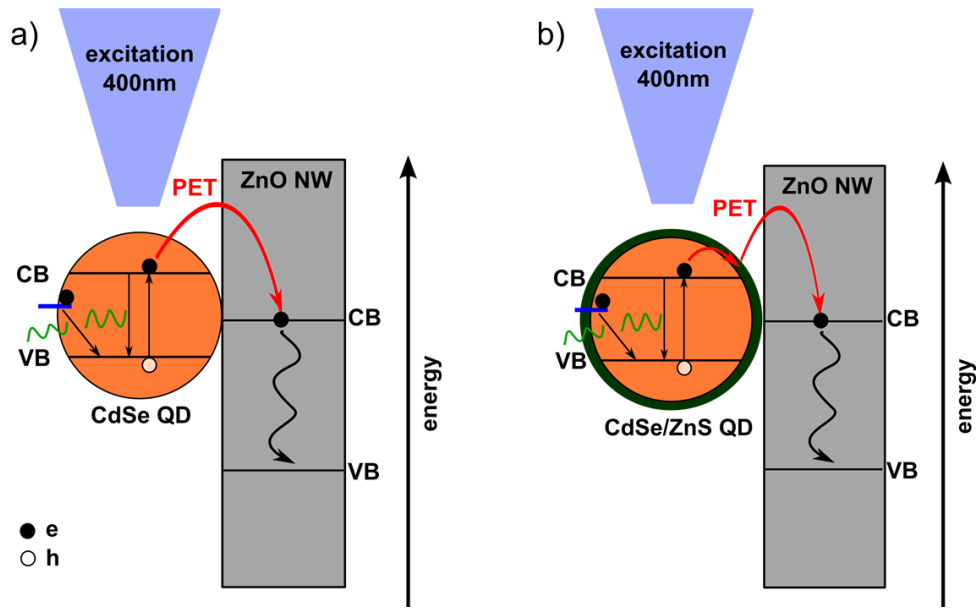

**Figure 6.** Schematic sketch of the photo-induced electron transfer (PET) and of the exciton recombination processes. a) PET from the CdSe QD into the ZnO nanowire. b) Core/shell QDs show electron tunneling into the shell and subsequent PET.

To describe the relaxation and recombination processes inside the quantum dots and the tunneling process into the ZnO NWs (schematic sketches of the processes, see Figure 6), a rate equation model was used. Figure 7 shows the energy diagram of the electron relaxation and recombination paths within the QD/NW system.

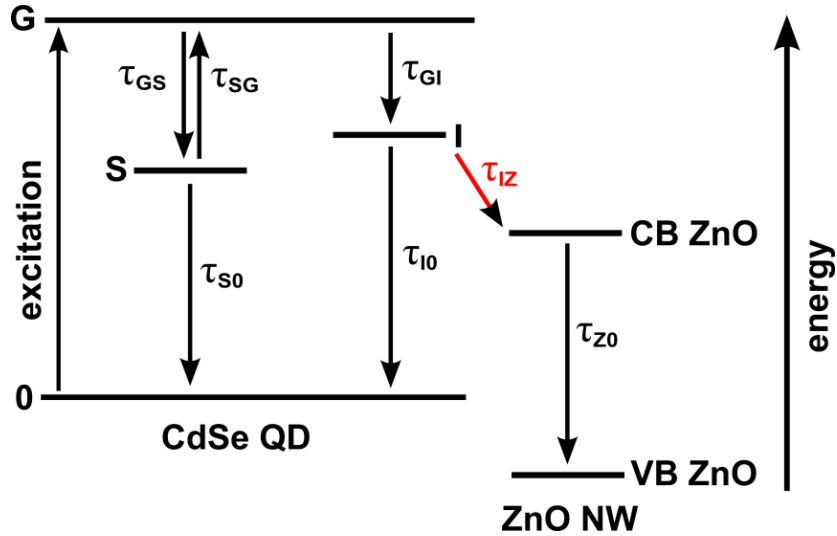

**Figure 7.** Schematic energy diagram of the exciton relaxation and recombination paths for the quantum dots, i.e., direct recombination from the excited state G via (indices GI, I0), surface recombination (index GS, including thermal back excitation SG), and electron transfer into the ZnO NW (index IZ).

Electrons are excited from ground state 0 into state G. From there, they can relax with a time constant $\tau_{GI}$ into state I, the lowest excited energy state of the QD. From state I, the electrons recombine with a time constant $\tau_{I0}$ into the ground state by emitting photons. Surface states on the QD are taken into account by introducing the trapping time $\tau_{GS}$ for transitions from G to the surface trap level S. From state S, electrons can return into the ground state 0 with a time constant $\tau_{S0}$, or are thermally re-excited into state G with a time constant $\tau_{SG}$. If the QDs are connected to the ZnO nanowire surface, the electrons can alternatively tunnel from state I with a time constant $\tau_{IZ}$ into the conduction band of the ZnO NW. All processes that eventually remove the electrons from the conduction band of the wire (non-radiative traps, surface states, interaction with the QD systems and the linker molecules) are described by the time constant $\tau_{Z0}$. Taking all these processes into account, we end up with the following rate equation system:

$$(1) \quad \frac{dN_G(t)}{dt} = -\frac{N_G(t)}{\tau_{GI}}(M_I - N_I(t)) - \frac{N_G(t)}{\tau_{GS}}(M_S - N_S) + \frac{N_S(t)}{\tau_{SG}}$$

$$(2) \quad \frac{dN_I(t)}{dt} = +\frac{N_G(t)}{\tau_{GI}}(M_I - N_I(t)) - \frac{N_I(t)}{\tau_{I0}} - \frac{N_I(t)}{\tau_{IZ}}(M_Z - N_Z)$$

$$(3) \quad \frac{dN_S(t)}{dt} = +\frac{N_G(t)}{\tau_{GI}}(M_S - N_S(t)) - \frac{N_S(t)}{\tau_{S0}} - \frac{N_S(t)}{\tau_{SG}}$$

$$(4) \quad \frac{dN_Z(t)}{dt} = +\frac{N_I(t)}{\tau_{IZ}}(M_Z - N_Z(t)) - \frac{N_Z(t)}{\tau_{Z0}}$$

$$(5) \quad \frac{dN_0(t)}{dt} = +\frac{N_I(t)}{\tau_{I0}}$$

$N_{0,G,S,I,Z}(t)$ describes the number of electrons in the levels 0, G, S, I and Z (the latter corresponding to CB ZnO), respectively, at a time t. $M_{G,S,I,Z}$ corresponds to the maximum number of states in G, S, I, and Z, respectively. Following the Pauli principle, the number of non-occupied states in the excited levels I, S, and Z of the QD ensemble is then given by $(M_I - N_I(t))$, $(M_S - N_S(t))$, and $(M_Z - N_Z(t))$.

The time-dependent PL intensity is calculated from

$$I(t + \Delta t) \propto N_I(t + \Delta t) \propto N_0(t + \Delta t) - N_0(t).$$

It is assumed that every electron produces a photon when returning into state 0. Therefore, $N_0(t + \Delta t) - N_0(t)$ corresponds to the number of emitted photons per time interval $\Delta t$.

In Figure 8, the experimental PL transients are plotted together with the PL signal calculated with the rate-equation model. The simulations have been obtained by performing a numerical optimization using the classical Runge-Kutta method (*ode45, Matlab*) followed by a second optimization by the Nelder-Mead simplex method. Through variation of the optimized parameters, the routine tries to find the local minimum in the difference between the simulated and measured PL intensity.

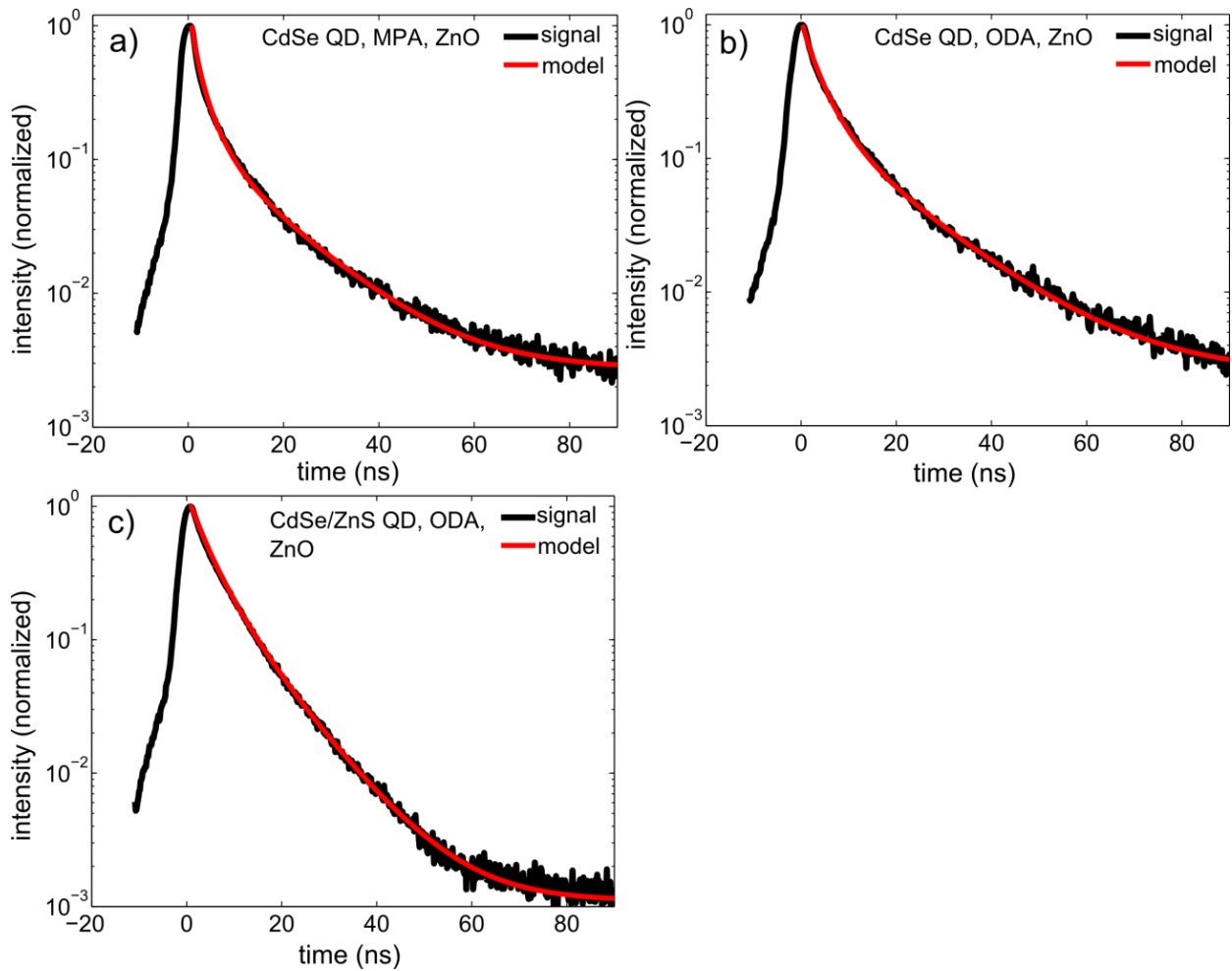

**Figure 8.** Normalized PL transients of the quantum dots attached to ZnO (black curves) and corresponding multi-exponential fits (red). a) MPA-capped QDs on ACG grown ZnO NWs, b) ODA-capped QDs on ACG ZnO NWs, c) ODA-capped core/shell QDs on ACG ZnO NWs.

In all three cases, the model is in excellent agreement with the experimental results over almost three orders of magnitude for the PL intensity. This agreement confirms that the most essential relaxation, recombination and tunneling pathways for the photo-excited electron-hole pairs are accounted for in the rate-equation model.

The parameters $\tau_{IZ}$ and $\tau_{Z0}$ from the optimized simulated transients in Figure 8 are shown in Table 3.

**Table 3.** Parameters $\tau_{IZ}$ and $\tau_{Z0}$ of QDs with different surface functionalization attached to ACG ZnO nanowires. $\tau_{GI}$, $\tau_{I0}$ are fixed to 18 and 6.8 ns for all cases, respectively (see text).

| sample | $\tau_{IZ}$ (ns) | $\tau_{Z0}$ (ns) |
|---|---|---|
| QD, MPA, ZnO | 1.6 | 862 |
| QD, ODA, ZnO | 5.7 | 1121 |
| core/shell QD, ODA, ZnO | 7.2 | 1088 |

By assuming that the QDs with different linker molecules have similar optical characteristics, what is in part justified by the absorption and emission characteristics given in Table 2., $\tau_{GI}$ and $\tau_{I0}$ are taken as dot-specific but fixed time constants. In combination with the model-optimized fixed starting values for $N_G$, $N_I$, $N_{ZnO}$ and the fixed saturation values $M_I$ and $M_{ZnO}$, respectively, the resulting fits of the experimental transients yield strongly differing $\tau_{IZ}$ for the three cases of MPA-capped CdSe QDs, ODA-capped CdSe QDs, and ODA-capped CdSe/ZnS core-shell QDs. $\tau_{IZ}$ is the crucial parameter governing the efficiency of the PET process. As the experimental results already revealed, $\tau_{IZ}$ is increasing with increasing molecule length. In case of the core/shell QDs the electrons are predominantly located at the interface between shell and QD core. Therefore, fast electron transfer is even more strongly hampered resulting in the slowest transient of all.

As mentioned before, all processes that eventually remove electrons from the conduction band of the ZnO nanowire are described by the time constant $\tau_{Z0}$. The variation of $\tau_{Z0}$ results from the model-optimized fitting in the region t > 100 ns. Different values of $\tau_{Z0}$ might be due to the fact that removal of electrons from the NW conduction band into surface states is reduced when these get passivated by longer linker molecules. Since the signal-to-noise ratio in this region is too low to extract significant data, and due to the fact that $\tau_{Z0}$ is very large compared to the other time constants, this time constant can almost be neglected for the interpretation of variations in $\tau_{IZ}$.

In Fig. 9, the significance of the parameter $\tau_{IZ}$ for fitting the measurement of MPA-capped QDs on ZnO becomes essentially visible. The sensitivity of the fit curves with regard to $\tau_{IZ}$ results from the strong influence of the linker molecule length and of the shell on the tunneling efficiency.

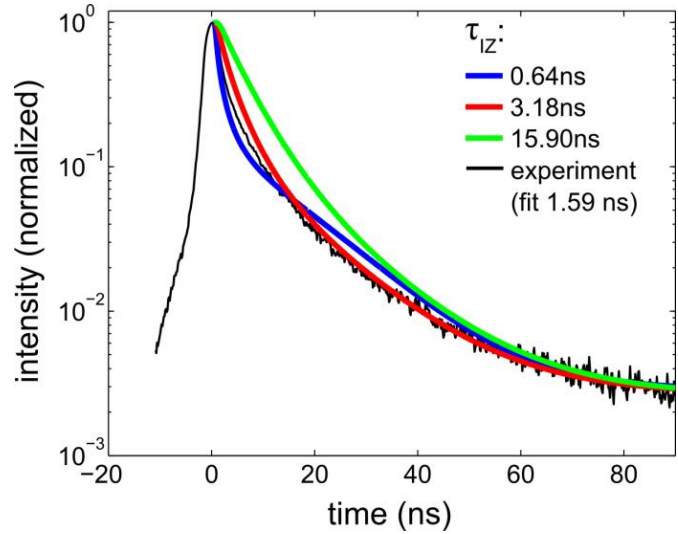

**Figure 9.** PL transient of MPA-capped QDs on ZnO (black) in comparison to fits (colors) for different $\tau_{IZ}$.

Small changes of $\tau_{IZ}$ result in strong variations in the PL transient fits. For long tunneling times $\tau_{IZ}$, i.e., less efficient tunneling, the number of electrons in the state I would increase. Thus, the transient becomes much slower than the experimentally observed one (compare green to black curve). Such a slow decay is rather more corresponding to the PL transient shown in Figure 8c for 'CdSe/ZnS, ODA, ZnO'. A very low value of $\tau_{IZ} \leq 1$ ns corresponding to a high tunneling rate into the ZnO and, consequently, to a rapid depopulation of the level I, yields a faster starting of the decay than experimentally observed.

## 4. Results of photoconductivity measurements

To confirm the electron transfer between the QDs and the ZnO nanowires to be of essential importance and to further understand how different linker molecules and oxidation processes influence the PET, the dynamics of the photocurrent of hybrid CdSe QD/ZnO NW devices was measured under selective excitation of the QDs (Figure 10).

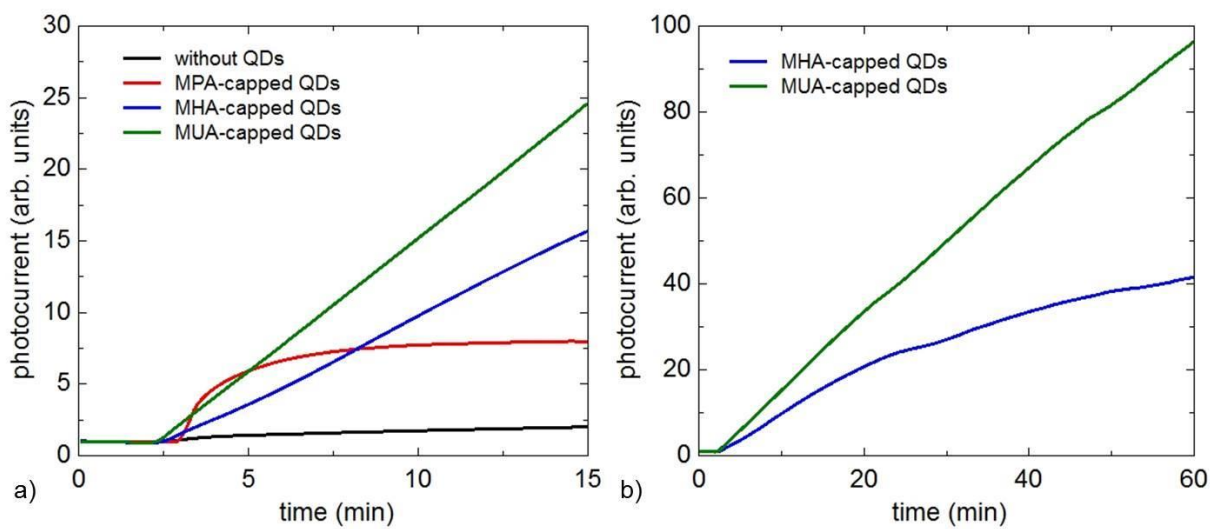

**Figure 10.** a) Photocurrent as a function of time for the first 15 minutes under resonant excitation into the QD transition. Black curve: NW excitation without QDs; colors: NW with QDs attached by different linker molecules. b) Photocurrent for MHA- and MUA-capped QDs over 60 minutes.

In Figure 10 a), the photocurrent response as a function of time is displayed. Prior to laser excitation, the original dark current value was measured for two minutes. The excitation energy of 2.69 eV is below the band gap of ZnO (3.37 eV at room temperature). The black curve is obtained for pure, uncoated, horizontally arranged NWs. It does not show a significant increase of photocurrent. For the NWs functionalized with MPA-capped QDs (red curve), the current increases quickly and reaches saturation after 5 minutes. For longer linker molecule lengths the photocurrent gets much higher but the increase becomes much slower, and saturation is not reached within a time of up to 15 minutes. The increase of the photocurrent for MUA-capped QDs is faster than for MHA-capped QDs. Saturation of the former is hardly visible for times even up to 60 minutes whereas the latter shows a trend to saturation (see Figure 10 b).

To summarize, we see the fastest increase of the photocurrent for the shortest molecule and the slowest for the longest one. However, the magnitude of the photocurrent is larger for longer linker molecules.

## 5. Discussion

The results presented in the previous section demonstrate the importance of electron tunneling from the QDs to the horizontal nanowires by the substantial increase of the photocurrent for photon energies far below the band gap energy of ZnO. Furthermore, a variation of the linker molecules, in particular of their length, drastically changes the photocurrent enhancement and its dynamics[20]. As shown in previous studies[20,21,22,23], desorption and readsorption of oxygen molecules from and on the oxide as well as the QD surfaces play an important role for the electron transfer processes between both material systems. Because of oxygen adsorption on the QD surface, surface defect states are passivated, and electrons from the QDs can efficiently tunnel into the conduction band of the ZnO NWs by PET. The holes remaining in the QDs can now capture electrons from the ZnO surface and thereby release adsorbed oxygen molecules from the NW surface. This results in a reduction of the upward band bending at the NW surface and effectively transfers electrons from bound surface states of the NW into free conduction band states, all processes mediated by the optical excitation of the QDs[22]. Saturation of the photocurrent is reached through a dynamic equilibrium between oxygen desorption and readsorption from and on both the nanowire and QDs surfaces.

The results gained from the photocurrent behavior according to Figure 10 therefore reveal that oxygen molecules can leave the nanowire surface faster if shorter linker molecules are used. However, the equilibrium between oxygen de- and adsorption at the surface will eventually be shifted to lower amounts of surface-adsorbed oxygen if longer linker molecules are present. This can be understood by considering that the linker molecules form a barrier layer on the NW and QD surfaces through which the oxygen molecules have to diffuse in order to leave the surfaces. Once they have passed this barrier, the probability for readsorption is substantially smaller for barriers formed by longer linker molecules because those are expected to the thicker. On the other hand, the diffusion through the thicker barriers will take more time than through thinner barriers as formed by the shorter linker molecules. Therefore, we can expect the PC dynamics to be faster for shorter linker molecules (smaller diffusion barriers for oxygen molecules) but the overall increase of the PC to be smaller (higher probability of oxygen readsorption).

Whereas the PC dynamics takes place on a time-scale of minutes, the PL dynamics presented in Figure 8 shows a time scale of a few tens of nanoseconds. Oxygen-related de- and adsorption processes together with molecule diffusion can therefore be ruled out to be of any significance for the relevant processes in this case. The comparison of the experimental results with the simulations based on the rate equation model (see Figure 7) reveals that the electron tunneling process from the lowest excited QD state (I) into the conduction band of the NW has a substantial impact on the PL dynamics. Compared with previous studies of hot electron tunneling[24] from QDs to metal oxide substrates of other material systems, the tunneling rate in our case is substantially slower (about a factor of 1,000). This is related to the fact that our experimental situation is not sensitive to hot electron tunneling processes. Here, tunneling processes of electrons dominate which are already relaxed to the lowest excited state of the QD and therefore have lost a large part of their excitation energy, mainly to phonons. The time-resolved PL measurements confirm that the tunneling rate is highest for the smallest spatial separation of the QDs from the NW surface (shortest linker molecules). Accordingly, the rate-equation model developed in section 3 assuming tunneling from the lowest excited QD state into the NW conduction band allows a very successful fit to the experimentally observed transients thus yielding a verification of our assumptions. In addition, the tunneling rate is further reduced when going from pure CdSe QDs to the core-shell CdSe/ZnS geometry (compare the experiments with identical linker molecules). In this case, the large band gap of the ZnS shell acts as an additional tunneling barrier for the photoexcited electrons and therefore substantially slows down the electron-transfer dynamics.

## 6. Conclusion

In conclusion, we have studied electron tunneling from CdSe QDs to ZnO NWs after selective photoexcitation of the QDs by time-resolved PL and photoconductivity experiments. We have shown that the PL dynamics is substantially influenced by tunneling of electrons from the lowest excited state of the QDs into the conduction band of the ZnO nanowires with a time constant $\tau_{IZ}$ ~ 1.6 – 7.2 ns depending on the length of the linker molecules between the QDs and the NW surface. Faster tunneling rates are achieved for shorter linker molecules, and an additional ZnS passivation layer on the CdSe quantum dots (core-shell structures) significantly reduces the tunneling rate. We find the tunneling rate from the lowest excited state of the QDs to be about three orders of magnitude smaller than that of typical hot-electron tunneling processes for other material systems, indicating that hot electrons do not play a major role in our case. We also demonstrate that the length of the linker molecules between the QDs and the NW surface has a substantial impact on the photoconductivity dynamics. If we chose a photon energy below the ZnO band-gap energy but large enough for absorption in the QDs, a faster increase of the PC signal is found for shorter linker molecules. The magnitude of the so obtained PC, however, is largest for the longest linker molecules. Assuming that the linker molecules constitute diffusion barriers for oxygen molecules whereby the thickness of the barrier scales with the molecule length we can explain the experimental results based on the typical model for PC in bare or functionalized ZnO nanowire samples.

## Acknowledgments

The authors acknowledge funding by the DFG research group FOR 1616 "Dynamics and Interactions of Semiconductor Nanowires for Optoelectronics". They thank Dongchao Hou (formerly Bremen) for having performed the synthesis of the MPA-capped CdSe QDs.

## Corresponding Author

*E-mail: bley@ifp.uni-bremen.de, Tel.: +49 (0)421 218 62207

address: Institute of Solid State Physics, University of Bremen, Otto-Hahn-Allee 1, 28359 Bremen, Germany

## Author Contributions

The manuscript was written through contributions of all authors. All authors have given approval to the final version of the manuscript.

Table of Contents Graphics:

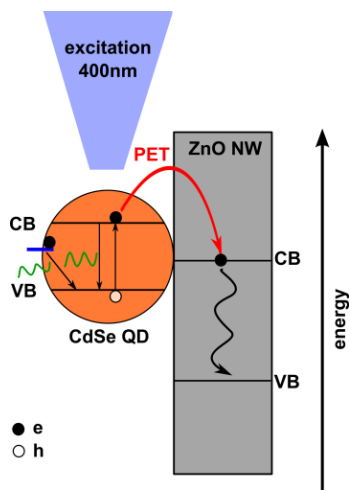